\documentclass[preprint,authoryear,12pt]{elsarticle}

\usepackage{algorithmic,algorithm}
\usepackage{amsmath,amssymb,amsthm}
\usepackage{color,graphicx}
\usepackage{array}
\usepackage{float,subfloat}
\usepackage{algorithmic,algorithm}
\usepackage{tikz,tikz-qtree}
\usepackage{multirow}
\usepackage{epstopdf}
\usepackage{setspace}
\usepackage[tight,TABTOPCAP]{subfigure}
\usepackage[textsize=tiny, textwidth=2.2cm]{todonotes}
\usepackage{bbold}
\usepackage{gensymb}

\DeclareMathOperator*{\argmax}{arg\,max}
\DeclareMathOperator*{\argmin}{arg\,min}

\def\x{{\mathbf x}}

\journal{Medical Image Analysis}

\begin{document}

\begin{frontmatter}

\title{Subject-Specific Abnormal Region Detection in Traumatic Brain Injury Using Sparse Model Selection on High Dimensional Diffusion Data}


\author[mymainaddress,mysecondaryaddress]{Matineh Shaker\corref{mycorrespondingauthor}}
\cortext[mycorrespondingauthor]{Corresponding author}

\author[mymainaddress]{Deniz Erdogmus}
\author[mymainaddress]{Jennifer Dy}
\author[mysecondaryaddress]{Sylvain Bouix}

\address[mymainaddress]{Electrical and Computer Engineering Department, Northeastern University, Boston, MA. shaker@ece.neu.edu,~erdogmus@ece.neu.edu,~jdy@ece.neu.edu }
\address[mysecondaryaddress]{Brigham and Women's Hospital, Harvard Medical School, Boston, MA. sylvain@bwh.harvard.edu  }


\begin{abstract}
We present a method to estimate a multivariate Gaussian distribution of diffusion tensor features in a set of brain regions based on a small sample of healthy individuals, and use this distribution to identify imaging abnormalities in subjects with mild traumatic brain injury. 
The multivariate model receives {\em apriori} knowledge in the form of a neighborhood graph imposed on the precision matrix, which models brain region interactions, and an additional $L_1$ sparsity constraint.  
The model is then estimated using the graphical LASSO algorithm and the Mahalanobis distance of healthy and TBI subjects to the distribution mean is used to evaluate the discriminatory power of the model. 
Our experiments show that the addition of the {\em apriori} neighborhood graph results in significant improvements in classification performance compared to a model which does not take into account the brain region interactions or one which uses a fully connected prior graph.
In addition, we describe a method, using our model, to detect the regions that contribute the most to the overall abnormality of the DTI profile of a subject's brain.
\end{abstract}

\begin{keyword}
\texttt{sparse learning} \sep \texttt{graphical lasso} \sep \texttt{TBI} \sep \texttt{DTI}
\end{keyword}

\end{frontmatter}

\section{Introduction}  \label{intro}

Abnormalities of Diffusion Tensor Imaging (DTI) data in neuroimaging studies are traditionally detected at the population level by directly comparing regions of interest across patients and healthy controls, and verifying whether distributions are statistically different in these regions. 
The assumption behind these types of analyses is that conditions in patients have homogeneous spatial patterns of abnormalities. 
However, in diseases such as traumatic brain injury (TBI) or multiple sclerosis, a common spatial pattern of injury is unlikely to occur, violating the main hypothesis of standard population studies. 

With an estimated 10 million people world-wide affected annually by a TBI, the burden that this condition imposes on society makes it a considerable public health problem \cite{hyder_impact_2007,feigin_incidence_2013,marion_proceedings_2011}. Importantly, a significant percentage (10-15\%) of individuals diagnosed with mild TBI experience persistent post-concussive symptoms (PPCS), which may lead to long-term disabilities \cite{bigler_neuropsychology_2008}. 
Symptoms range from physical, such as headache; cognitive, such as difficulty concentrating; and emotional/behavioral, such as irritability and impulsivity. 
In the majority of these chronic cases, there is no radiological evidence of injury from conventional magnetic resonance imaging (MRI) or computed tomography (CT), and little is known about the pathophysiology underlying the injury. 
Thus establishing radiological evidence of brain injury is a critical first step towards the proper diagnosis and monitoring of TBI, and may lead to establishing neuroimaging biomarkers to help predict recovery versus PPCS and to assess better the impact of therapies on the injured brain. 

Recent methods for injury detection in mild TBI patients have been developed by estimating a model of "healthy" DTI features and testing whether brain regions have outside-of-normal-range values for a particular subject's brain (see \cite{mayer2014methods} for a nice overview).
Typically, each region is modeled by the mean and standard deviation of the DTI feature of interest over all healthy individuals, and individual TBI subject's data are z-transformed using these healthy population parameters.
Finally regions with a z-score above a given threshold (typically 2 standard deviations) are flagged as abnormal, and statistics such as the number of abnormal regions or the average z-score over the brain are compared between TBI and controls.
Methods mostly differ from each other based on how the mean and standard deviation are estimated, and how bias is avoided when testing normal controls that have been used to estimate the "healthy" model parameters \cite{ge2005applications,kim2013whole,bouix2013increased}.
Most methods study one DTI feature at a time (except for \cite{hellyer2013individual}, which uses four DTI features in a multivariate setting), but none of the current techniques model the inter-dependence of DTI features between neighboring brain regions.
Another interesting result from our previous work, suggest that DTI changes are observable in gray matter regions in these patients (potentially related to glial scaring), and thus one should study the full brain as opposed to only white matter in this population \cite{bouix2013increased}.

In this paper, we extend the multiple univariate setting of \cite{bouix2013increased} to a high dimensional Gaussian multivariate model which accounts for inter-region interactions. 
One of the main challenge we need to overcome is a relatively small number of healthy subjects (in the order of 50) compared to the number of parameters to estimate (in the order of 10,000).
Our method thus relies on the estimation of a sparse representation of the region co-dependencies as modeled by a precision matrix. 

Although not as thoroughly studied in diffusion MRI, sparse representation of inter-region interactions is the subject of much research in fMRI. 
Extracted networks capture higher order dependencies among variables, and therefore are effective in exploring local interactions of brain regions \cite{friston2011functional}.
Unfortunately, the estimation of these functional connectivities from subject to subject can be difficult to do robustly and recent research has focused on imposing a prior to the sparse representation. 
One such example is the work of \cite{zhu2013exploring}, which uses structurally-weighted least absolute shrinkage and selection operator (LASSO) regression, and models the directional functional interactions of resting state fMRI data based on {\em structural} connectivity constraints encoded by 358 cortical landmarks derived from DTI data \cite{zhu2012dicccol}. 

Our work is similar in spirit, with some key differences. 
Here, we use DTI to evaluate subtle tissue changes in TBI patients by detection of outliers compared to a model of normal brain derived from 145 brain regions of 34 healthy subjects. 
A feature vector containing fractional anisotropy (FA) measures over 145 brain regions represents each subject. 
We model the distribution of these features in the healthy subjects as a multi-dimensional Gaussian distribution as represented by a precision matrix. 
Our method relies on the theorem that conditional independence of two variables given others is equivalent to setting the corresponding precision matrix entity to zero \cite{lauritzen1996graphical}.
We leverage this theorem by imposing a brain neighborhood prior graph on the structure of the precision matrix, reducing the number of parameters to estimate by favoring interactions of proximal regions and ignoring the interactions of regions which are far away from each other.\footnote{Note that a sparse precision matrix does not imply a sparse covariance matrix; therefore distant brain regions are not assumed to be independent with this constraint -- only conditionally independent as shown in Eq. (\ref{eq:theorem}).} 
The multi-dimensional Gaussian model is further regularized by an $L_1$ sparsity constraint and estimated using the graphical LASSO \cite{friedman2008sparse}.

\section{Gaussian graphical models}
Let $\mathbf{x} = [X_1, X_2, .., X_d]$ be a $d$-dimensional random vector so that it has a multivariate Gaussian distribution $\mathbf{x} \sim \mathcal{N}(\mu,\Sigma)$, with $d$-dimensional mean vector $\mu$, and a $d \times d$ covariance matrix $\Sigma$. 
In a Gaussian graphical model, an unweighted undirected graph with adjacency matrix $G$, can be used to represent the conditional dependence structure between the individual variables $X_i$. 
More specifically, the edge structure of $G$ can be imposed onto the inverse covariance matrix, also known as the precision matrix, $\Sigma^{-1} \equiv \Theta = \{ \theta_{ij}\}$, and {\em conditional independence} between $X_i$ and $X_j$ can be expressed as a zero in the corresponding location in $\Theta$:
    \begin{equation}
        \label{eq:theorem}    
        X_i \perp\!\!\!\perp X_j  
        \Leftrightarrow  
        G_{ij}=0
        \Leftrightarrow  
        \theta_{ij}=0
    \end{equation} 
The proof can be found in \cite{lauritzen1996graphical}. 

One key benefit of this representation is that one can use a priori information to impose a conditional independence structure to the model.
This is particularly useful in scenarios where a high dimensional $\Theta$ needs to be estimated with only a few samples, and expert knowledge about the data set can help guide sparse model learning. 
By using a graph $G$ which sets many of the precision matrix elements to zero before the estimation process, we can greatly reduce the number of parameters of the model, and thereby increase the robustness of the optimization.

In addition, we assume global sparsity of the model, and thus add an $L_1$ penalty term to further regularize the model. Following \cite{banerjee2008model}, let $X$ be the $n \times d$ data matrix representing $n$ observations, $S$ be the $d \times d$ sample covariance matrix, and $G$ the a priori graph, the maximum a-posteriori (MAP) estimate of $\Theta$ given X and G is:


    \begin{eqnarray}
        \label{eq:map}    
        \hat{\Theta}_{MAP} &=& \argmax_{\Theta} \log{p(\Theta|X,G)} \nonumber \\
        &=& \log{\det{\Theta}}-tr{(S\Theta)}-\rho \left\|\Theta\right \|_{L_1} \\
    & & \text{with } \theta_{ij}=0 \text{ when } G_{ij}=0. \nonumber
    \end{eqnarray}
\noindent where $\rho$ is a scalar controlling the $L_{1}$ norm penalty weight.

The optimization method uses the graphical LASSO algorithm \cite{friedman2008sparse}, which can elegantly incorporate $G$ into the optimization process.

In the following section, we describe how this graphical model with the addition of an a priori graph can be applied to the problem of estimating a multivariate Gaussian distribution of DTI features in healthy subjects and use this model to detect brain injuries in subjects with mild TBI.

\section{Application to injury detection in TBI}
Our driving hypothesis for using graphical models is that brain regions next to each other have similar, or at least highly related, DTI signal in healthy subjects.
We thus model these interactions by only considering edges connecting proximal regions in the graph imposed on the precision matrix.
If a TBI subject has a region with abnormal signal, having modeled the healthy region-to-region interaction will help us increase our sensitivity to classifying a TBI brain as abnormal, compared to looking at each region independently.

\subsection{Subjects and data acquisition}
In this work, we used the data described in \cite{bouix2013increased}. 
There are $n=34$ healthy subjects, $p=11$ TBI patients who reported symptoms (see Table \ref{tab:subjects} for details), such as headaches, emotional dysregulation and memory impairments at the time of data collection, as well as $m=11$ normal controls demographically matched to TBIs. 
The normal controls are separated from healthy subjects for validation purposes.
\begin{table}[!ht]
\caption{Description of individual mTBI subjects adapted from \cite{bouix2013increased}}
\label{tab:subjects}
\scriptsize
\centering
\begin{tabular}{|c|c|c|c|c|l|}
\hline
\hline
ID   & Age & Gender & Source    & Duration     & Symptoms                        \\
     &     &        & of Injury & since Injury &                                 \\ 
\hline
TB01 & 45  & F      & MVA*      &  17.0        & Cognitive impairment, emotional \\
     &     &        &           &              & dysregulation, depression       \\
\hline
TB02 & 38  & M      & MVA       & 106.6        & Mild memory impairment, mild    \\   
     &     &        &           &              & executive function impairment   \\
     &     &        &           &              & emotional dysregulation         \\
\hline
TB03 & 44  & F      & MVA       & 121.3        & Dizziness, exhaustion, periodic \\
     &     &        &           &              & limb movements, hypersomnia     \\
     &     &        &           &              & depression and anxiety          \\
\hline
TB04 & 30  & M      & Sports    & 2.6          & Diplopia, fatigues easily,      \\
     &     &        & Injury    &              & executive function impairment   \\
\hline
TB05 & 42  & M      & MVA       & 138.0        & Cognitive impairment, memory    \\
     &     &        &           &              & executive function impairment   \\
\hline
TB06 & 28  & M      & Assault   & 27.0         & Anxiety, depression, insomnia   \\
     &     &        &           &              & ADHD, intrusive thoughts,       \\
     &     &        &           &              & memory deficits, overeating     \\
\hline
TB07 & 24  & M      & Blast     & 70.3         & Anxiety, dpanic attacks,        \\
     &     &        & Exposure  &              & hypervigilance, overeating,     \\
     &     &        &           &              & difficulty concentrating        \\
\hline
TB08 & 25  & M      & Blast     & 83.3         & Depression, memory impairment   \\
     &     &        & Exposure  &              & difficulty w/rapidly presented  \\
     &     &        &           &              & information                     \\
\hline
TB09 & 29  & M      & Blast     & 51.4         & Irritability, nightmares,       \\
     &     &        & Exposure  &              & depression, panic attacks,      \\
     &     &        &           &              & cognitive and memory impairment \\
\hline
TB10 & 24  & M      & Blast     & 55.9         & Headaches, memory impairment,   \\
     &     &        & Exposure  &              & problems concentrating,         \\
     &     &        &           &              & irritability, anxiety, nightmares\\
\hline
TB11 & 39  & M      & Sports    & 9.5          & Facial pain, memory/executive   \\
     &     &        & Injury    &              & function impaired, emotional    \\
     &     &        &           &              & dysregulation                   \\
\hline
\end{tabular}
\\
*MVA= Motor Vehicle Accident
\end{table}
Subjects underwent MRI scanning, including a high resolution diffusion tensor imaging scan and a high resolution structural T1 weighted scan. 
Each T1 image was segmented using the FreeSurfer software \cite{fischl2002whole}, resulting in 176 gray matter (GM), white matter (WM), and cerebrospinal fluid (CSF) sections. 
CSF sections and sections smaller than 300 $mm^3$ were excluded from the analysis, as these smaller regions led to unstable estimation of mean/std of the DTI measures and many failed to pass normality tests.
The remaining 145 sections (83 in GM and 62 in WM) were registered onto the diffusion space using a non-linear diffeomorphic registration algorithm~\cite{avants2011registration}. 
The average FA was computed in each region for each subject.
The outcome of the image processing procedure is a feature vector of the average FA in $d=145$ brain structures in each subject. 
More details about data acquisition and processing can be found in \cite{bouix2013increased}.
In addition, the same procedure was applied for the other standard DTI measures: mean diffusivity (MD), radial diffusivity (RD), and axial diffusity (AD).

\subsection{a priori graph} \label{sec:expertprior}
Given this data set, we will have to estimate a $145 \times 145$ precision matrix based on 34 observations.
In order to reduce the number of parameters to estimate, we chose to design a simple graph that will only consider the relationship between neighboring regions in the brain.
Two regions were considered to be neighbors if they were connected in a template FreeSurfer segmentation using 26-connectivity. 
Our motivation for choosing this graph for TBI stems from the knowledge that nearby regions in healthy subjects will tend to have similar tissue properties and thus similar DTI signal (note that we are not considering tensor orientation).

We have made the choice of connecting neighboring GM and WM regions with an edge as there is increasing evidence that the tissue and geometric properties of proximal GM/WM regions are stongly related \cite{miyata2009schiz,koch2013cortex,liu2014MRI,savadjiev2014media}.
Furthermore, the graph is only a “guide” for the precision matrix estimation process. If the data does not support the existence of a (conditional) relationship between two variables, the corresponding entry in the precision matrix will converge to zero even if it was linked by an edge in the prior graph.

\begin{figure}[!ht]
        \centering
        \subfigure[Regions IDs in a brain slice]{
            \includegraphics[width=.4\textwidth]{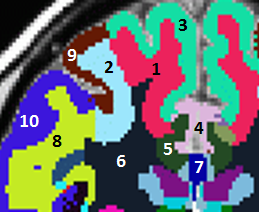}
            \label{fig:brainslice}
        } 
        \subfigure[Illustration of a neighborhood graph corresponding to the brain regions in (a). Bold lines show the subgraph associated with region 1. ]{
            \includegraphics[width=.4\textwidth]{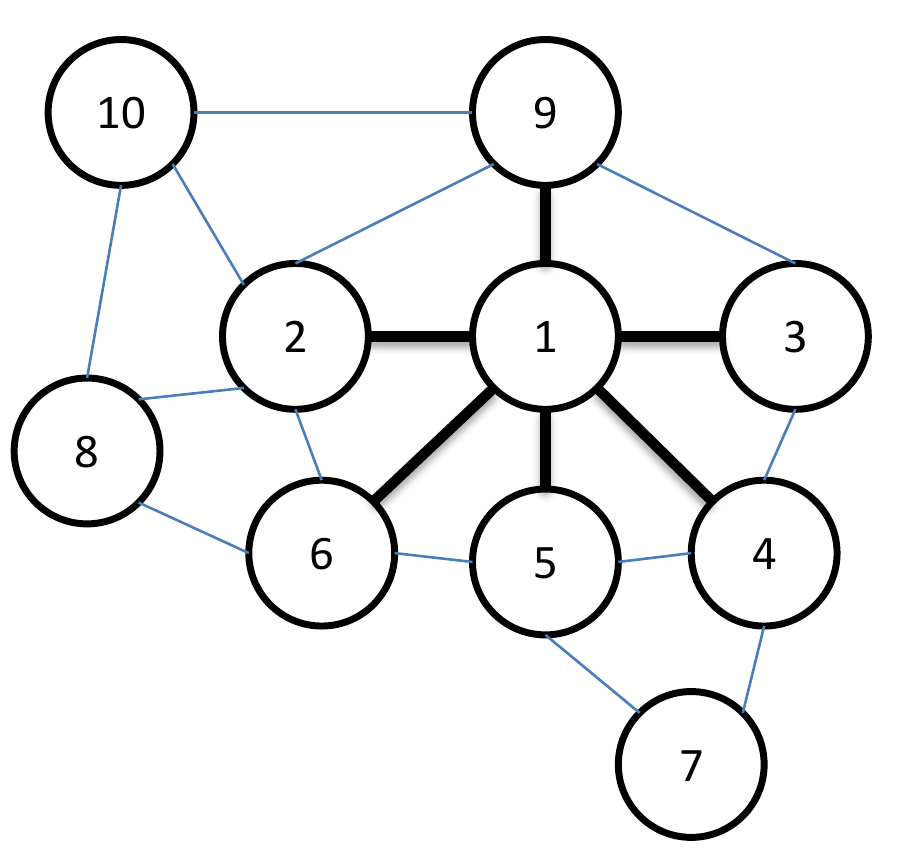}
            \label{fig:subgraph}
        }
        \subfigure[Neighborhood graph adjacency matrix for the full brain]{
            \includegraphics[width=.4\textwidth]{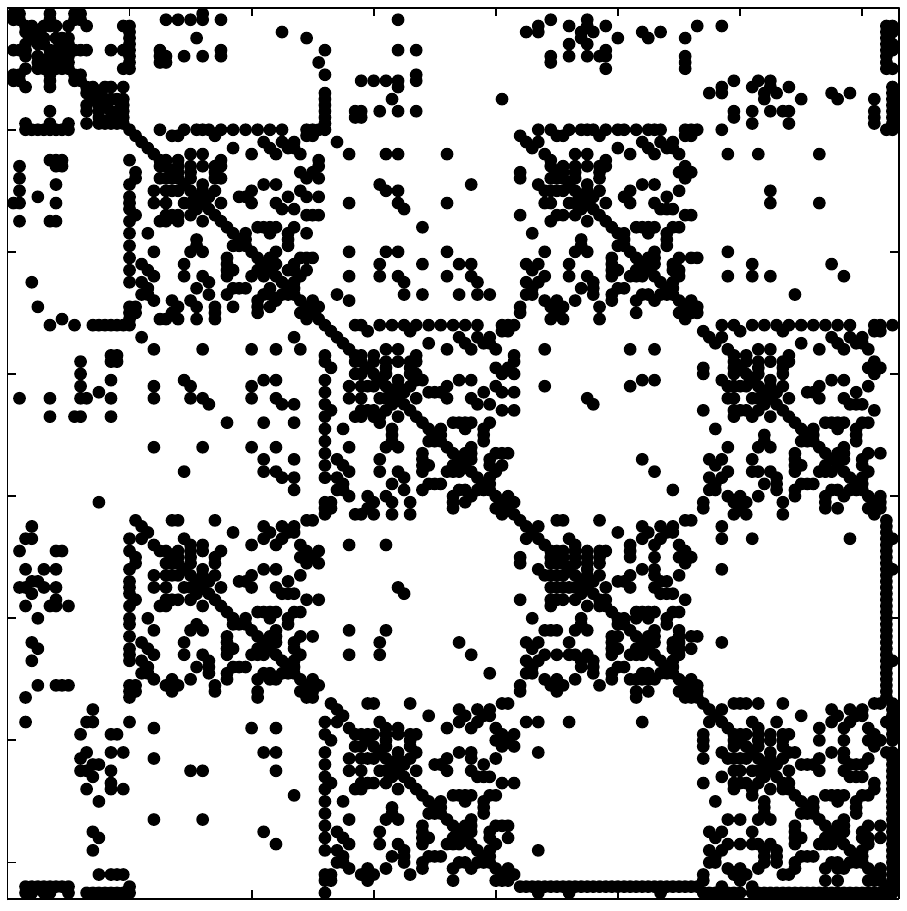}
            \label{fig:adjacencymatrix}
        }
        \caption{Illustration of the prior graph through 10 brain regions. 
        The vector assigned to region 1 in the adjacency matrix is 
        $[1 1 1 0 1 1 0 0 0 0]$, emphasizing connections of regions 
        2, 3, 5, 6 and disconnections of regions 4, 7, 8, 9, 10 to region 1. Bold lines in Figure~\ref{fig:subgraph} show the subgraph associated with region 1.}
        \label{fig:graph}
    \end{figure}

The neighborhood network $G$ is illustrated in Figure~\ref{fig:graph}. 
Each brain structure is represented as a node in the graph, and conditional dependence is only considered between regions connected by an edge, whereas all other relationships are ignored. 
Bold lines in Figure~\ref{fig:subgraph} show the subgraph associated with region 1.
The adjacency matrix corresponding to the complete neighborhood graph is shown in Figure~\ref{fig:adjacencymatrix}. One can observe a large number of parameters that will be set to 0 in $\Theta$. 
Note that the conditional independence of two non-neighboring regions imposed by this graph does not enforce unconditional independence; pairs of regions that are not immediate neighbors are allowed to have correlations.


\subsection{Identifying an abnormal brain}
Let $X$ be the $n \times d$ matrix representing the set of $d$ features in $n$ healthy subjects, $Y$ the $m \times d$ matrix capturing the observations in $m$ normal controls, and $Z$ the $p \times d$ matrix representing the set of $p$ TBI patients. 
Normal controls are healthy subjects matched to patients demographically, and are separated from the healthy training set $X$ for validation purposes. 

The overall design is to generate a model $(\mu_X, \Theta_X)$ based on the healthy subject data $X$ and test whether a TBI subject $i$ is abnormal by measuring the Mahanobis distance of its feature vector $\mathbf{z_i}$ to the model:
   \begin{equation}
        \label{eq:mahalanobis}    
        d_M(\mathbf{z_i}) = \sqrt{(\mathbf{z_i}-\mu_X)^T \Theta_X (\mathbf{z_i}-\mu_X)}
    \end{equation} 

As the Mahalanobis distance follows a $\chi^2$ distribution, a threshold for an abnormal brain based on this distance can be theoretically derived (e.g., above the 95th percentile of the expected Mahalanobis distances). 
However, in our work, we test the discriminatory power of our model by computing the Mahalanobis distances of TBI subjects ($Z$) and matched controls ($Y$) and evaluate its classification performance using Receiver Operating Characteristic (ROC) curve analysis.

\subsection{Identifying individual abnormal regions}
\label{sec:detection}
The method we have presented thus far has the ability to identify whether a subject's imaging profile is {\em overall} abnormal.
The natural next step is to identify which regions are most affected in this subject and thus provide some information that could potentially be linked to the pathophysiology of the brain injury, or help targeting therapies to particular brain areas.
Given $k$ regions, we propose a greedy forward sorting approach to identify these abnormal regions as follows.
Let $R_s$ be the ordered set of sorted regions from most normal to most abnormal, $R_u = \{1,..,k\}$ be the set of all regions, and $d_R$ be the Mahalanobis distance computed by only taking into account the regions in subset $R \subset R_u$.
We build $R_s$ by incrementally adding the region $r_i \in R_u \setminus R_s$, which minimizes $d_{R_i}$, where $R_i=R_s \cup \{r_i\}$. 
This process is repeated until all regions have been sorted from most normal to most abnormal.
The procedure is detailed in Alg. \ref{alg:sorting}

\begin{algorithm}
\caption{Sorting regions from most normal to most abnormal}
\label{alg:sorting}
    \begin{algorithmic}[1]
      \STATE $R_u = \{1,..,k\}$
      \STATE $R_s = ()$
      \FOR {$i$: 1 to k}
         \STATE $r_i = \argmin_{j \in R_u \setminus R_s} (d_{R_s \cup \{j\}})$
         \STATE $R_s = R_s \cup \{r_i\}$
      \ENDFOR
      \RETURN $R_s$
    \end{algorithmic}
\end{algorithm}

\begin{figure}[!ht]
\centering
\subfigure[Region sorting and selection in TBIs]{
    \includegraphics[width=.45\textwidth]{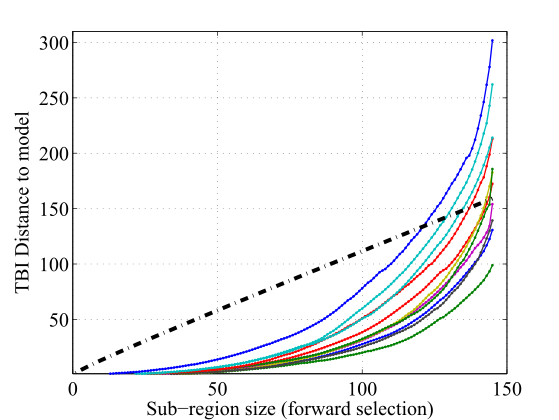}
    \label{fig:tbi_fwd}
}
\subfigure[Region sorting and selection in controls]{
    \includegraphics[width=0.45\textwidth]{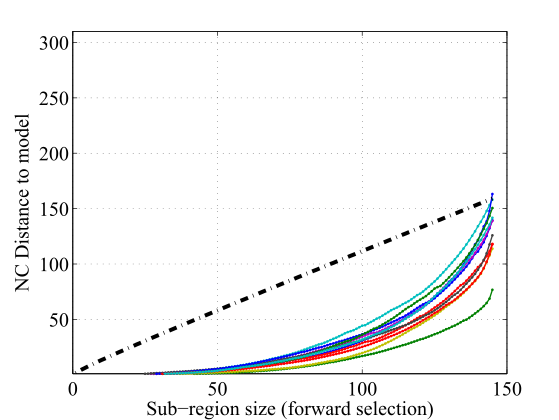}
    \label{fig:nc_fwd}
}
\caption{Abnormal Region detection based on the $\chi^2$ distribution and our greedy forward sorting of regions. The dashed lines correspond to the 95th percentile of the CDF of the $\chi^2$ distribution. Each of the colored curves represent, for each subject, the accumulated Mahalanobis distance given by the subsets of sorted regions. Regions added after the point at which the subjects Mahalanobis distance curve exceeds the 95\% threshold on the CDF of $\chi^2$ distribution are flagged as abnormal.
}
\label{fig:greedy_selection}
\end{figure}

The output of this algorithm is an ordering of regions along with $k$ Mahalanobis distances, $d_{R_i}$ of the corresponding subsets of sorted regions. 
The last step consists of comparing the subject's sorted $D_i$s with the theoretical distribution of the Mahalanobis distance (the  $\chi^2$ distribution with $i$ degrees of freedom) and finding the first region after which the subject's sorted distances exceed the 95th percentile of the $\chi^2$ distribution.
Let $F_{\chi^2}(D,l)$ be the cumulative distribution function of the $\chi^2$ distribution with $l$ degrees of freedom and $\hat{k} = \argmax_{k} ( F_{\chi^2}(D_{k}, k ) < 0.95 )$.
Thanks to our sorting process, the regions that are not in the subset of size $\hat{k}$ will generate increasingly unlikely Mahalanobis distances and can be flagged as abnormal.
This thresholding procedure is illustrated in Figure \ref{fig:greedy_selection}.

\section{Experiments}   \label{sec:experiments}
In order to evaluate the performance of the prior neighborhood graph approach, we tested three different graph structures as follows:
    \begin{enumerate}
        \item The neighborhood prior graph as described in Section~\ref{sec:expertprior}, with 
        an $L_1$ sparsity constraint.
        \item A node-only graph with all off-diagonal elements set to zero in the precision matrix.
        \item A fully connected graph evaluating all off-diagonal terms with an $L_1$ sparsity constraint.
    \end{enumerate}

We tested the robustness of each model by performing a cross validation procedure as follows.
Using a leave-one-out strategy, we generated $n-1$ models $(\mu_i,\tilde{\Theta_i})$ from $X_{|i}$, the set of healthy subjects $X$ without the $i$-th element. 
For each model  $(\mu_i,\tilde{\Theta_i})$, we then calculate $d_{M,X_{|i}}$ for all TBI subjects in $Z$ and for all control subjects in $Y$.
In addition, we repeated this procedure for a range of regularization parameter $\rho$ (from $10^{-2}$ to $10$) to evaluate the impact of this parameter on performance. 
Thus, for each $\rho$ we had $n$ sets of ``TBI vs. Controls'' Mahalanobis distances and were able to compute confidence intervals of various classification performance measures (in our case the area under the receiver operating characteristic curve -- AUC). 

As described earlier, the maximization of the posterior distribution in (\ref{eq:map}), iteratively minimizes certain edges of the graph in two ways: 1) Data driven, where natural interaction of variables among all samples estimate the edges in the graph or precision matrix elements; 2) Prior model driven, where a predefined graph is imposed to the model which sets certain edges to zero, without iterative learning.

In the following experiments, the performance of the node only graph (diagonal precision matrix) is evaluated to illustrate the importance of multivariate vs. univariate analyses. Graphical LASSO is clearly not needed in this diagonal precision matrix design.

\subsection{Node-only versus neighborhood versus fully-connected graphs}

\begin{figure}[!ht]
        \centering
        \subfigure[AUC as a function of $\rho$]{
            \includegraphics[width=.45\textwidth]{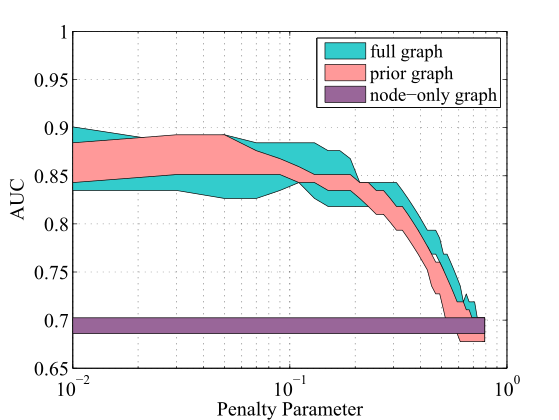}
            \label{fig:conf_intervals_auc}
        }
        \subfigure[BIC as a function of $\rho$]{
            \includegraphics[width=.45\textwidth]{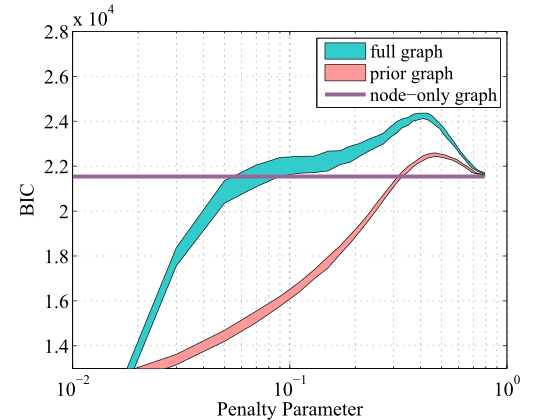}
            \label{fig:conf_intervals_bic}
        }
        \subfigure[Complexity of models]{
            \includegraphics[width=.45\textwidth]{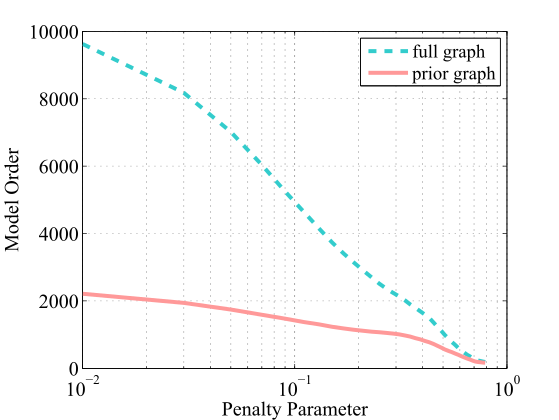}
            \label{fig:model_orders}
        }
        \subfigure[Performance vs. model complexity]{
            \includegraphics[width=.45\textwidth]{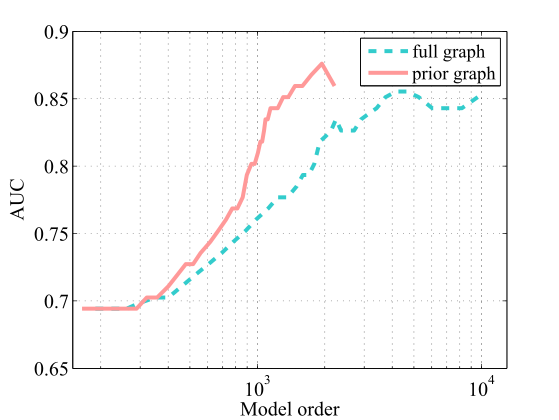}
            \label{fig:auc_vs_modelorder}
        }
        \caption{(a) 90\% CIs of the AUC, as a function of the penalty parameter $\rho$ for the prior graph, fully-connected graph, and node-only graph; (b) 90\% CIs of the BIC as a function of the penalty parameter $\rho$ for each model; (c) Number of model parameters for a given penalty weight for prior and fully-connected graphs; (d) AUC as a function of the number of model parameters for the prior and fully-connected graphs. Both models show similar AUC performance, but the prior graph has better model complexity and BIC.}
        \label{fig:cross_val_results}
    \end{figure}

In Figure~\ref{fig:cross_val_results}, all three graph types are examined.
In addition, the evaluation is performed for different $\rho$ values to observe the impact of this regularization parameter on the classifier performance. 

Figure~\ref{fig:conf_intervals_auc} compares the 90\% confidence intervals (CI) of the AUC($\rho$) functions of 34 cross-validation instances across graph types. 
The confidence intervals are computed using the functional box plot method \cite{sun2011functional}, and the envelope of the 90\% central region is shown in Figure~\ref{fig:cross_val_results}. 
One can observe that both the neighborhood and the full graph clearly outperform the node-only model. 
In addition, while these two graphs have comparable average performance over all cross-validation, the neighborhood graph has a tighter 90\% confidence interval.

The advantage of the prior graph over a fully-connected graph is even clearer when considering the Bayesian information criterion (BIC), as given by
    \begin{equation}
    \label{eq:bic}
        BIC = -2 \: \ln{p(X|\hat{\Theta})} + j \: \ln{n}, 
    \end{equation}
\noindent where $p(X|\hat{\Theta})$ is the maximized value of the likelihood function, $j$ is the number of parameters estimated, and $n$ is the number of training samples.
In our case, $j$ represents the number of non-zero values in the estimated precision matrix $\hat{\Theta}$. 
BIC is a criterion for model selection among a finite set of models, and balances the goodness of fit ($p(X|\hat{\Theta})$) with a penalty term for the number of model parameters.
This criterion penalizes models which increase their likelihood by overfitting the data.
Using BIC as a model selection criteria, the prior graph model is preferred due to its lower BIC. 
Figure~\ref{fig:model_orders} compares the number of parameters estimated (model order) for the two multivariate models. 
In Figure~\ref{fig:auc_vs_modelorder}, one can observe that the neighborhood graph always has a higher AUC than the full graph for the same model complexity.  


\subsection{Neighborhood versus random prior graphs}
\begin{figure}[!ht]
        \centering
        \subfigure[AUC central regions for random graphs]{
            \includegraphics[width=.47\textwidth]{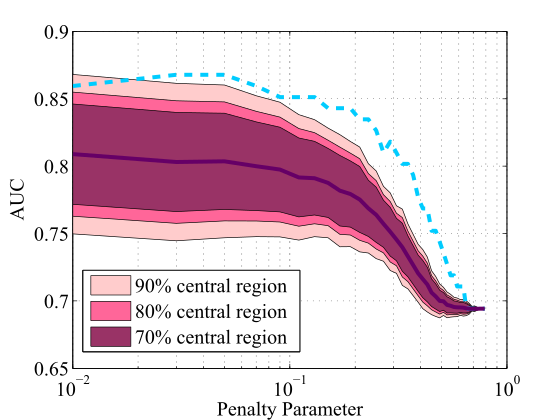}
            \label{fig:random_auc}
        }
        \subfigure[BIC central regions for random graphs]{
            \includegraphics[width=.47\textwidth]{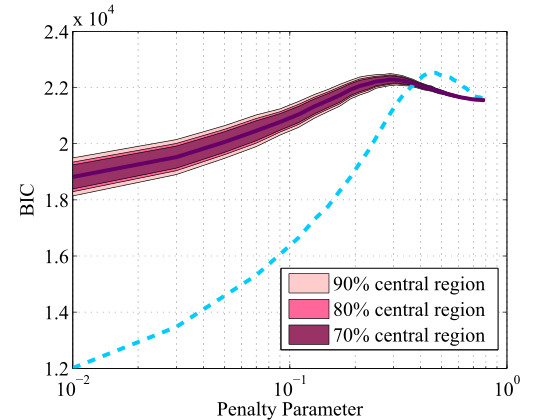}
            \label{fig:random_bic}
        }
        \caption{Comparison of the neighborhood graph model performance (blue dashed line) with 1000 random graphs with the same number of edges.}
        \label{fig:random}
    \end{figure}

To check the importance of expert knowledge in model selection, 1000 random graphs were generated so that they have the same number of edges as the prior graph but at uniformly random locations. 
Figure~\ref{fig:random} compares the AUC and BIC of the neighborhood graph to the 75\%, 85\% and 95\% central regions of the randomly generated graphs. 
The neighborhood graph has an AUC that is higher than the 95\% central region of random graphs. 
Similarly the BIC is almost always lower for the neighborhood graph than it is for random graphs.
This result illustrates that the better performance of the neighborhood prior is not due to overfitting, but because of the selection of an appropriate graphical model. 
The percentile of the prior graph performance at various $\rho$ compared to the random graphs distribution is shown in Table~(\ref{tab:percentiles}).
    


    \begin{table}[!ht]
        \caption{percentile of the prior graph performance on the random graph distribution}
        \label{tab:percentiles}
        \scriptsize
        \centering
        \begin{tabular}{|c|c|c|c|c|c|c|c|}
            \hline
            $\rho$        & 0.01    & 0.02    & 0.1    & 0.2    & 0.4   & 0.8   \\
            \hline
            Percentile    & 98\% & 99\% & 100\% & 98\% & 100\% & 100\% \\
            \hline
        \end{tabular}
    \end{table}

\subsection{Selecting the optimal penalty parameter $\rho$}

\begin{figure}[!ht]
    \centering
        \includegraphics[width=.5\textwidth]{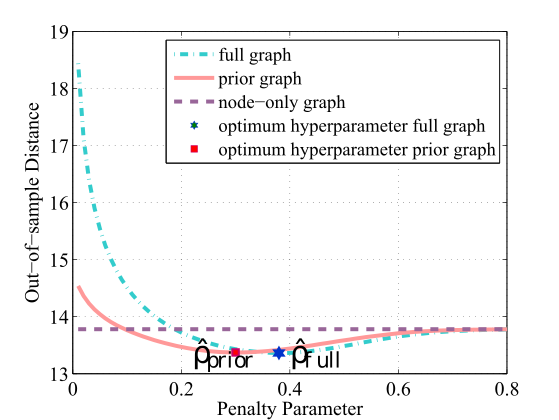}
    \caption{Selection of the optimum penalty parameter for each model based on minimizing  leave-one-out sum of squared Mahalanobis distances to the model mean.}
    \label{fig:opt_rho}
\end{figure}

While the above analyses provide valuable information on the quality of the different models under different regularization by the parameter $\rho$, one does need to select a single optimal $\hat{\rho}$ value to estimate the final model. 
In order to find this optimum, the Mahalanobis distance of each training point to the model mean estimated with the remaining training points is calculated. 
The optimum $\rho$ minimizes the leave-one-out sum of squared distances, which is $\hat{\rho}=0.3$ for the prior graph and $\hat{\rho}=0.38$ for the full graph, as shown in Figure~\ref{fig:opt_rho}. 
Table~\ref{tab:roc_results} compares the performance of the three models at optimum values of $\rho$. 
Once again, the neighborhood prior graph model outperforms both the node-only and full graph priors.
Note that the performance of the node-only graph does not depend on the value of $\rho$. 

In order to put our results in context with traditional "z-score" approaches \cite{white2009white,lipton2012robust,bouix2013increased,mayer2014methods}, we also performed the computation of the AUC of the mean absolute z-score over all regions as a potential measure to distinguish patients from controls.

Z-scores were computed with respect to the mean and standard deviation of FA in each region over $X$, the training set of healthy subjects.
As expected, this method does not perform as well as the multivariate models.

\begin{table}[!ht]
\caption{AUC analysis for optimal multivariate models and independent z-scores}
\label{tab:roc_results}
\scriptsize
\centering
\begin{tabular}{| c || c | c | c | c |}
\hline
\hline
& Full Graph & Prior Graph & Node-only Graph & mean absolute zscore  \\
\hline

\textbf{AUC} & 0.83 & 0.86 & 0.69 & 0.65\\
\hline

\textbf{Sensitivity} & 0.64 & 0.73 & 0.73 & 0.64\\
\hline
\textbf{Specificity} & 0.91 & 1 & 0.64 & 0.64\\
\hline

\end{tabular}
\end{table}

\subsection{Investigating other DTI measures}
The z-score analysis of \cite{bouix2013increased} only found statistically significant differences for FA. Nevertheless, we further tested, using our multivariate method, the other most common DTI measures: Mean Diffusivity (MD), Axial Diffusivity (AD), and Radial Diffusivity (RD). 
For all experiments, we used the prior graph and the same regularization parameter $\hat{\rho}=0.3$.
As in the previous work, only FA reached significance, although we hypothesize that AD could reach significance given a larger sample size (see Table \ref{tab:dti_results}).
Consequently, all subsequent analyses focused solely on FA.

\begin{table}[!ht]
\caption{p-value of Wilcoxon ranksum tests and AUC for the most common DTI metrics.}
\label{tab:dti_results}
\scriptsize
\centering
\begin{tabular}{| c || c | c |}
\hline
\hline
Measure     & p     & AUC \\
\hline
\textbf{FA} & 0.016 & 0.86\\
\textbf{MD} & 0.168 & 0.68\\
\textbf{AD} & 0.088 & 0.72\\
\textbf{RD} & 0.265 & 0.64\\
\hline
\end{tabular}
\end{table}

\subsection{Correlations with behavioral measures}
Similarly to \cite{bouix2013increased}, we performed Spearman correlations between the Mahalanobis distance and behavioral measures in BI subjects. 
The results presented in Table \ref{tab:corr_results} are very similar to our previous work, with "Digit Symbol", a measure of processing speed, the only behavioral test significantly correlated with imaging (rho=-0.62, p=0.04), although reported p-values are uncorrected for multiple comparisons. The Bonferroni corrected significance threshold is 0.004

Nevertheless, our sample of 11 TBI subjects is quite small, and we expect better correlations with a larger number of subjects. Confidence intervals on rho is calculated according to the formula presented in \cite{ruscio2008constructing}.

\begin{table}[!ht]
\caption{Spearman correlations between Behavioral measures and Mahalanobis distances based on FA in TBI subjects. Note that p-values are uncorrected for multiple comparisons, the Bonferroni corrected significance threshold is 0.004}
\label{tab:corr_results}
\scriptsize
\centering
\begin{tabular}{| l || l | c | c | c | c |}
\hline
\hline
Test              & Subtest                 & rho  & p (uncorrected) & $CI_{min}$ & $CI_{max}$ \\
\hline
California Verbal & Trials 1-5              &  0.27 & 0.42  & -0.40 & 0.75 \\
Learning Test II  & Short Delay Free Recall & -0.57 & 0.07  & -0.87 & 0.04 \\
                  & Short Delay Cued Recall & -0.37 & 0.26  & -0.80 & 0.30 \\
                  & Long Delay Free Recall  & -0.35 & 0.29  & -0.78 & 0.31 \\
                  & Long Delay Cued Recall  & -0.23 & 0.49  & -0.73 & 0.42 \\
\hline
Processing Speed  & Digit Symbol            & -0.62 & 0.04* & -0.89 &-0.04 \\
                  & Symbol Search           & -0.45 & 0.16  & -0.82 & 0.20 \\
\hline
Digit Span        & Digit Span              & -0.30 & 0.37  & -0.76 & 0.36 \\
\hline 
Trail Making      & Trail Making A          & -0.02 & 0.96  & -0.61 & 0.59 \\
                  & Trail Making B          &  0.37 & 0.26  & -0.30 & 0.80 \\
\hline 
Controlled Oral   &                         & -0.18 & 0.60  & -0.70 & 0.47 \\
Word Association  &                         &       &       &       &      \\
\hline 
STROOP            &                         &  0.17 & 0.61  & -0.47 & 0.70 \\
\hline
\end{tabular}
\end{table}

\subsection{Individual abnormal regions identification}
\begin{figure}[!ht]
    \centering
    \subfigure[]{
        \includegraphics[width=.47\textwidth]{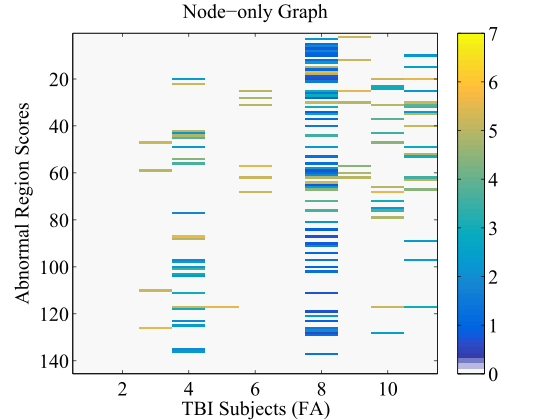}
        \label{fig:tbi_colormap_nodeonly}
    }
    \subfigure[]{
        \includegraphics[width=.47\textwidth]{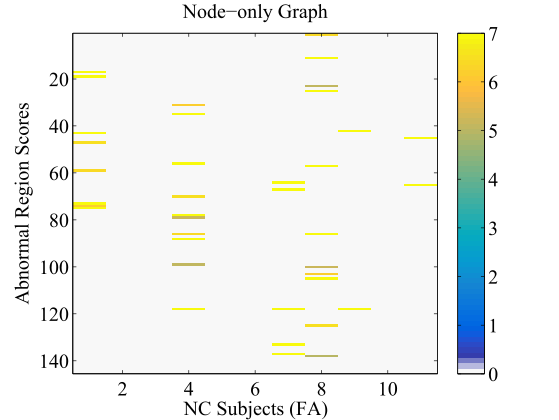}
        \label{fig:nc_colormap_nodeonly}
    }
    \subfigure[]{
        \includegraphics[width=.47\textwidth]{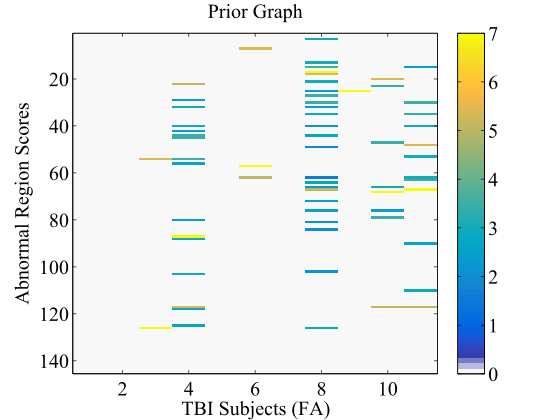}
        \label{fig:tbi_colormap_prior_inter}
    }
    \subfigure[]{
        \includegraphics[width=.47\textwidth]{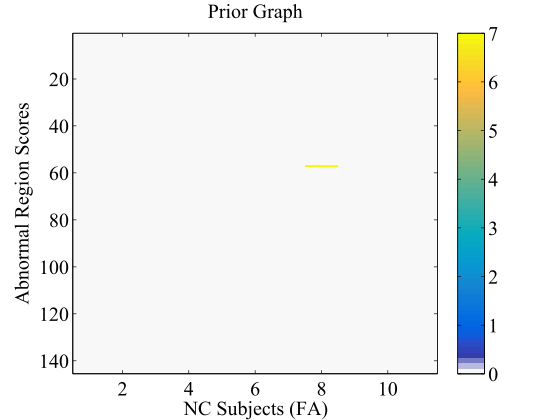}
        \label{fig:nc_colormap_prior_inter}
    }
    \subfigure[]{
        \includegraphics[width=.47\textwidth]{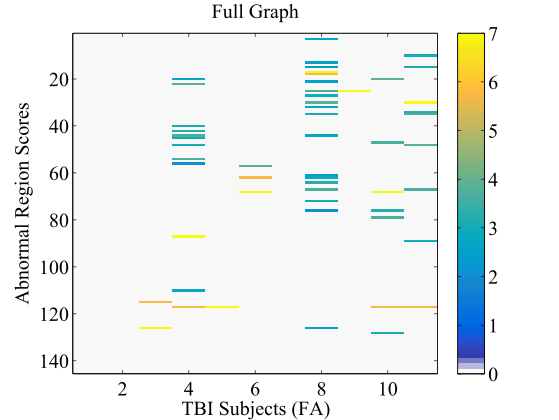}
        \label{fig:tbi_colormap_full_inter}
    }
    \subfigure[]{
        \includegraphics[width=.47\textwidth]{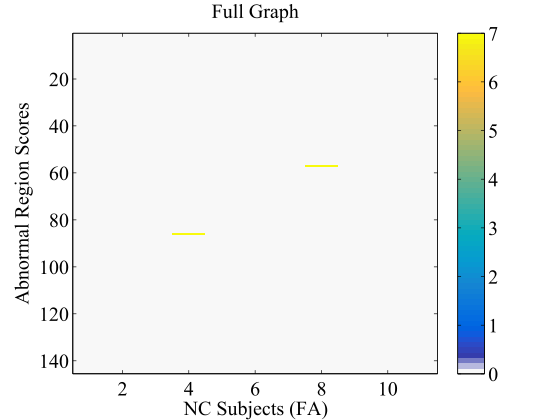}
        \label{fig:nc_colormap_full_inter}
    }
    \caption{Abnormality maps show which regions are affected in each subject for different graphical models. Left: TBI subjects, Right: Normal Controls (NC). Top to Bottom: node-only graph, neighborhood graph, full graph.}
    \label{fig:abnormal_regions}
\end{figure}

In this section, we present the results of the detection of individual abnormal regions as described in section \ref{sec:detection}.
Each subfigure in Figure \ref{fig:abnormal_regions} shows a $k \times l$ matrix. 
The $k$ rows represent the regions and the $l$ columns the individual subjects.
The intensity associated with each region in each figure corresponds to its respective amount of "abnormality".
We define this abnormality $a_i$ as the following differential \[a_i = \frac{(D_{i}-D_{i-1})}{(\tilde{D_{i}}-\tilde{D_{i-1}})},\] where $D_{i}$ is the Mahalanobis distance of the sorted subset of size $i$ and $\tilde{D}_i$ is 95\% threshold of the $\chi^2$ distribution, i.e., $F_{\chi^2}(\tilde{D_{i}}, i )=0.95$.

\begin{figure}[!ht]
    \centering
    \subfigure[]{
        \includegraphics[width=.47\textwidth]{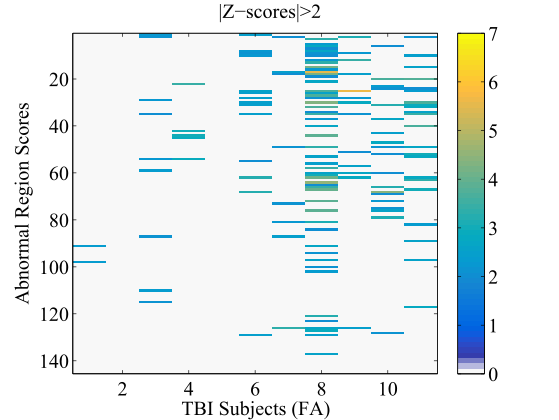}
        \label{fig:tbi_colormap_z2}
    }
    \subfigure[]{
        \includegraphics[width=.47\textwidth]{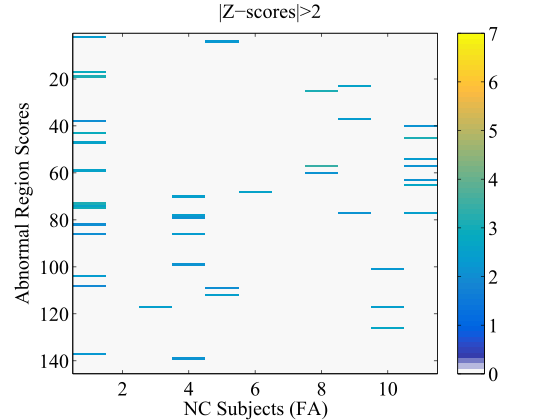}
        \label{fig:nc_colormap_z2}
    }
    \subfigure[]{
        \includegraphics[width=.47\textwidth]{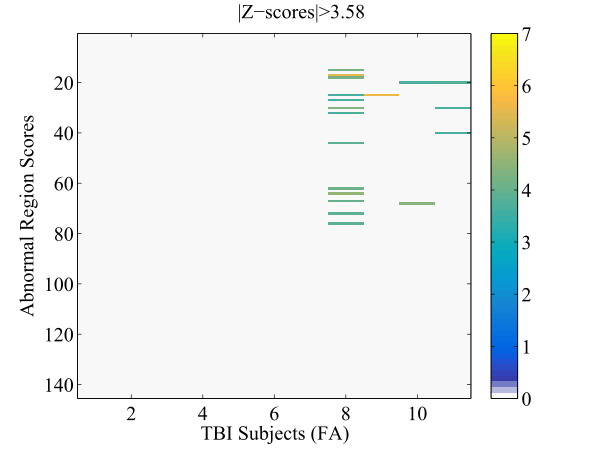}
        \label{fig:tbi_colormap_z3}
    }
    \subfigure[]{
        \includegraphics[width=.47\textwidth]{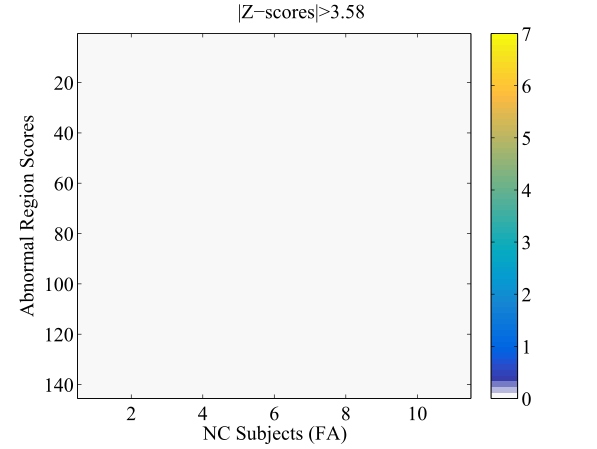}
        \label{fig:nc_colormap_z3}
    }
    \caption{Abnormality maps based on z-scores for two different thresholds. Top: threshold is $|z|>2$, Bottom: $|z|>3.58$ (Bonferroni correction for the number of regions)}
    \label{fig:abnormal_zscores}
\end{figure}

We also present the equivalent figures for standard z-score analyses in Figure \ref{fig:abnormal_zscores}.
In this figure, we present regions with an absolute z-score greater than 2 as well as those greater than 3.58, the threshold corresponsing to a Bonferroni correction for the number of regions.

One can observe that both the neighborhood and full graph display similar patterns of detections, whereas the node-only graph displays many false positives. 
The z-score method show similar results to the node only graph at $|z|>2$ and a subset of the multivariate techniques at $|z|>3.58$.

\section{Discussion}
Graphical models are a powerful and flexible technique to impose a structure on a multivariate Gaussian model, which has allowed us to constrain the estimation of a model of DTI signal based on a small data set of healthy subjects. 

We chose to constrain a LASSO estimation procedure of a precision matrix, by imposing a \textit{conditional independence} structure on our model \cite{lauritzen1996graphical}. 
Note the emphasis on {\em conditional} independence, i.e., the lack of an edge in our prior graph does not forbid covariance between two variables, but assumes that for two variables X \& Y, knowing X offers no additional information about Y given what we already know from the other variables in the model.  
Therefore, the independence structure imposed by the graph is quite flexible and allows for the examination of many relationships including those of regions that are very far apart.

We applied this method to detect whether subjects who experienced a TBI had an abnormal DTI scan, by measuring the Mahalanobis distance of their data to the model. 
We tested three different graph structures, a node-only graph, a fully connected graph, and a neighborhood graph, which only connects regions that are next to each other in the brain. 
The ability of each method to accurately detect an abnormal brain was tested by classifying TBI vs NC subjects using their Mahalanobis distance to the model under study and computing the corresponding AUC.

Our results demonstrate that multivariate approaches (full and neighborhood graph) clearly outperform the univariate approaches, inluding standard z-score analyses \cite{white2009white,lipton2012robust,bouix2013increased,mayer2014methods}. While both full and neighborhood graph show similar AUCs, the neighborhood graph leads to a better model when taking into account model complexity, i.e., the number of non-zero elements in the precision matrix.
Furthermore, our cross-validation experiments show that although the sample size is small, the results are quite robust as the 90\% central region width of the AUC is less than 0.05 for the neighborhood graph. 
Moreover, the neighborhood model always outperforms randomly generated graph with the same number of edges, indicating that the ``expert'' knowledge embedded in the graph is indeed a valuable prior to constrain the estimation of the model.

Importantly, the flexibility of graphical models can allow us to test a number of prior graphs, including network-based graph generated from diffusion MRI and/or functional MRI network analyses \cite{yoldemir2015ipmi,vergara2016neuro}.
This is certainly a topic we plan to investigate in future work.
Another possible extension is the study of DTI (or more generally diffusion MRI) measures in combination, by using a nested precision matrix design, although larger sample sizes would be needed for such complex models.
We are particularly interested in diffusion MRI measures related to neuroinflammation such as free water, as it may be a marker for subjects experiencing chronic symptoms \cite{pasternak2014neuro,planetta2016brain}

We have also shown that our multivariate analysis can detect individual regions with abnormal data.
In fact, our results show fewer false positives in NCs and more regions detected in mTBIs compared to classical independent z-score analyses. 
Nevertheless, this aspect of our work was exploratory and further development inspired by factor analysis techniques should be investigated.

Finally, we tested the connection between imaging data and symptomatology, but unfortunately were not able to find strong relationships between behavioral measures and DTI beyond a single measure (Digit Symbol, a measure of processing speed). 
We believe the main reason is the small sample size, but also the fact that we have only looked at the overall Mahalanobis distance (a global imaging measure). 
With more data, one could investigate connections between symptoms and subsets of regions corresponding to known networks associated with a particular brain function ((e.g., \cite{han2016neuro}), which we think will lead to stronger relationships between imaging and behavioral measures.


\subsection*{Acknowledgements} 
This work was supported in part by a CIMIT Soldier in Medicine Award; NSF grants CCF 1442728, IIS-1149570, and IIS-1118061; NIH grants R01 NS 078337 and RO1HL089856; DoD grants W81XWH-08-2-0159; and a Veterans Administration Merit Review Award.

\bibliographystyle{model2-names}
\bibliography{refs1}

\begin{thebibliography}{30}
\expandafter\ifx\csname natexlab\endcsname\relax\def\natexlab#1{#1}\fi
\providecommand{\url}[1]{\texttt{#1}}
\providecommand{\href}[2]{#2}
\providecommand{\path}[1]{#1}
\providecommand{\DOIprefix}{doi:}
\providecommand{\ArXivprefix}{arXiv:}
\providecommand{\URLprefix}{URL: }
\providecommand{\Pubmedprefix}{pmid:}
\providecommand{\doi}[1]{\href{http://dx.doi.org/#1}{\path{#1}}}
\providecommand{\Pubmed}[1]{\href{pmid:#1}{\path{#1}}}
\providecommand{\bibinfo}[2]{#2}
\ifx\xfnm\relax \def\xfnm[#1]{\unskip,\space#1}\fi
\bibitem[{Avants et~al.(2011)Avants, Tustison, Song, Cook, Klein and
  Gee}]{avants2011registration}
\bibinfo{author}{Avants, B.B.}, \bibinfo{author}{Tustison, N.J.},
  \bibinfo{author}{Song, G.}, \bibinfo{author}{Cook, P.A.},
  \bibinfo{author}{Klein, A.}, \bibinfo{author}{Gee, J.C.},
  \bibinfo{year}{2011}.
\newblock \bibinfo{title}{A reproducible evaluation of ants similarity metric
  performance in brain image registration.}
\newblock \bibinfo{journal}{NeuroImage} \bibinfo{volume}{54},
  \bibinfo{pages}{2033--2044}.
\bibitem[{Banerjee et~al.(2008)Banerjee, El~Ghaoui and
  d'Aspremont}]{banerjee2008model}
\bibinfo{author}{Banerjee, O.}, \bibinfo{author}{El~Ghaoui, L.},
  \bibinfo{author}{d'Aspremont, A.}, \bibinfo{year}{2008}.
\newblock \bibinfo{title}{Model selection through sparse maximum likelihood
  estimation for multivariate gaussian or binary data}.
\newblock \bibinfo{journal}{The Journal of Machine Learning Research}
  \bibinfo{volume}{9}, \bibinfo{pages}{485--516}.
\bibitem[{Bigler(2008)}]{bigler_neuropsychology_2008}
\bibinfo{author}{Bigler, E.D.}, \bibinfo{year}{2008}.
\newblock \bibinfo{title}{Neuropsychology and clinical neuroscience of
  persistent post-concussive syndrome}.
\newblock \bibinfo{journal}{J Int Neuropsychol Soc} \bibinfo{volume}{14},
  \bibinfo{pages}{1--22}.
\newblock \DOIprefix\doi{10.1017/S135561770808017X}.
\bibitem[{Bouix et~al.(2013)Bouix, Pasternak, Rathi, Pelavin, Zafonte and
  Shenton}]{bouix2013increased}
\bibinfo{author}{Bouix, S.}, \bibinfo{author}{Pasternak, O.},
  \bibinfo{author}{Rathi, Y.}, \bibinfo{author}{Pelavin, P.E.},
  \bibinfo{author}{Zafonte, R.}, \bibinfo{author}{Shenton, M.E.},
  \bibinfo{year}{2013}.
\newblock \bibinfo{title}{Increased gray matter diffusion anisotropy in
  patients with persistent post-concussive symptoms following mild traumatic
  brain injury}.
\newblock \bibinfo{journal}{PloS one} \bibinfo{volume}{8},
  \bibinfo{pages}{e66205}.
\bibitem[{Feigin et~al.(2013)Feigin, Theadom, Barker-Collo, Starkey, McPherson,
  Kahan, Dowell, Brown, Parag, Kydd, Jones, Jones, Ameratunga and {BIONIC Study
  Group}}]{feigin_incidence_2013}
\bibinfo{author}{Feigin, V.L.}, \bibinfo{author}{Theadom, A.},
  \bibinfo{author}{Barker-Collo, S.}, \bibinfo{author}{Starkey, N.J.},
  \bibinfo{author}{McPherson, K.}, \bibinfo{author}{Kahan, M.},
  \bibinfo{author}{Dowell, A.}, \bibinfo{author}{Brown, P.},
  \bibinfo{author}{Parag, V.}, \bibinfo{author}{Kydd, R.},
  \bibinfo{author}{Jones, K.}, \bibinfo{author}{Jones, A.},
  \bibinfo{author}{Ameratunga, S.}, \bibinfo{author}{{BIONIC Study Group}},
  \bibinfo{year}{2013}.
\newblock \bibinfo{title}{Incidence of traumatic brain injury in {New}
  {Zealand}: a population-based study}.
\newblock \bibinfo{journal}{Lancet Neurol} \bibinfo{volume}{12},
  \bibinfo{pages}{53--64}.
\newblock \DOIprefix\doi{10.1016/S1474-4422(12)70262-4}.
\bibitem[{Fischl et~al.(2002)Fischl, Salat, Busa, Albert, Dieterich,
  Haselgrove, Van Der~Kouwe, Killiany, Kennedy, Klaveness
  et~al.}]{fischl2002whole}
\bibinfo{author}{Fischl, B.}, \bibinfo{author}{Salat, D.H.},
  \bibinfo{author}{Busa, E.}, \bibinfo{author}{Albert, M.},
  \bibinfo{author}{Dieterich, M.}, \bibinfo{author}{Haselgrove, C.},
  \bibinfo{author}{Van Der~Kouwe, A.}, \bibinfo{author}{Killiany, R.},
  \bibinfo{author}{Kennedy, D.}, \bibinfo{author}{Klaveness, S.}, et~al.,
  \bibinfo{year}{2002}.
\newblock \bibinfo{title}{Whole brain segmentation: automated labeling of
  neuroanatomical structures in the human brain}.
\newblock \bibinfo{journal}{Neuron} \bibinfo{volume}{33},
  \bibinfo{pages}{341--355}.
\bibitem[{Friedman et~al.(2008)Friedman, Hastie and
  Tibshirani}]{friedman2008sparse}
\bibinfo{author}{Friedman, J.}, \bibinfo{author}{Hastie, T.},
  \bibinfo{author}{Tibshirani, R.}, \bibinfo{year}{2008}.
\newblock \bibinfo{title}{Sparse inverse covariance estimation with the
  graphical lasso}.
\newblock \bibinfo{journal}{Biostatistics} \bibinfo{volume}{9},
  \bibinfo{pages}{432--441}.
\bibitem[{Friston(2011)}]{friston2011functional}
\bibinfo{author}{Friston, K.J.}, \bibinfo{year}{2011}.
\newblock \bibinfo{title}{Functional and effective connectivity: a review}.
\newblock \bibinfo{journal}{Brain connectivity} \bibinfo{volume}{1},
  \bibinfo{pages}{13--36}.
\bibitem[{Ge et~al.(2005)Ge, Law and Grossman}]{ge2005applications}
\bibinfo{author}{Ge, Y.}, \bibinfo{author}{Law, M.}, \bibinfo{author}{Grossman,
  R.I.}, \bibinfo{year}{2005}.
\newblock \bibinfo{title}{Applications of diffusion tensor mr imaging in
  multiple sclerosis}.
\newblock \bibinfo{journal}{Annals of the New York Academy of Sciences}
  \bibinfo{volume}{1064}, \bibinfo{pages}{202--219}.
\bibitem[{Han et~al.(2016)Han, Chapman and Krawczyk}]{han2016neuro}
\bibinfo{author}{Han, K.}, \bibinfo{author}{Chapman, S.B.},
  \bibinfo{author}{Krawczyk, D.C.}, \bibinfo{year}{2016}.
\newblock \bibinfo{title}{{{D}isrupted {I}ntrinsic {C}onnectivity among
  {D}efault, {D}orsal {A}ttention, and {F}rontoparietal {C}ontrol {N}etworks in
  {I}ndividuals with {C}hronic {T}raumatic {B}rain {I}njury}}.
\newblock \bibinfo{journal}{J Int Neuropsychol Soc} \bibinfo{volume}{22},
  \bibinfo{pages}{263--279}.
\bibitem[{Hellyer et~al.(2013)Hellyer, Leech, Ham, Bonnelle and
  Sharp}]{hellyer2013individual}
\bibinfo{author}{Hellyer, P.J.}, \bibinfo{author}{Leech, R.},
  \bibinfo{author}{Ham, T.E.}, \bibinfo{author}{Bonnelle, V.},
  \bibinfo{author}{Sharp, D.J.}, \bibinfo{year}{2013}.
\newblock \bibinfo{title}{Individual prediction of white matter injury
  following traumatic brain injury}.
\newblock \bibinfo{journal}{Annals of neurology} \bibinfo{volume}{73},
  \bibinfo{pages}{489--499}.
\bibitem[{Hyder et~al.(2007)Hyder, Wunderlich, Puvanachandra, Gururaj and
  Kobusingye}]{hyder_impact_2007}
\bibinfo{author}{Hyder, A.A.}, \bibinfo{author}{Wunderlich, C.A.},
  \bibinfo{author}{Puvanachandra, P.}, \bibinfo{author}{Gururaj, G.},
  \bibinfo{author}{Kobusingye, O.C.}, \bibinfo{year}{2007}.
\newblock \bibinfo{title}{The impact of traumatic brain injuries: a global
  perspective}.
\newblock \bibinfo{journal}{NeuroRehabilitation} \bibinfo{volume}{22},
  \bibinfo{pages}{341--353}.
\bibitem[{Kim et~al.(2013)Kim, Branch, Kim and Lipton}]{kim2013whole}
\bibinfo{author}{Kim, N.}, \bibinfo{author}{Branch, C.A.},
  \bibinfo{author}{Kim, M.}, \bibinfo{author}{Lipton, M.L.},
  \bibinfo{year}{2013}.
\newblock \bibinfo{title}{Whole brain approaches for identification of
  microstructural abnormalities in individual patients: comparison of
  techniques applied to mild traumatic brain injury}.
\newblock \bibinfo{journal}{PloS one} \bibinfo{volume}{8},
  \bibinfo{pages}{e59382}.
\bibitem[{Koch et~al.(2013)Koch, Schultz, Wagner, Schachtzabel, Reichenbach,
  Sauer and Schlosser}]{koch2013cortex}
\bibinfo{author}{Koch, K.}, \bibinfo{author}{Schultz, C.C.},
  \bibinfo{author}{Wagner, G.}, \bibinfo{author}{Schachtzabel, C.},
  \bibinfo{author}{Reichenbach, J.R.}, \bibinfo{author}{Sauer, H.},
  \bibinfo{author}{Schlosser, R.G.}, \bibinfo{year}{2013}.
\newblock \bibinfo{title}{{{D}isrupted white matter connectivity is associated
  with reduced cortical thickness in the cingulate cortex in schizophrenia}}.
\newblock \bibinfo{journal}{Cortex} \bibinfo{volume}{49},
  \bibinfo{pages}{722--729}.
\bibitem[{Lauritzen(1996)}]{lauritzen1996graphical}
\bibinfo{author}{Lauritzen, S.L.}, \bibinfo{year}{1996}.
\newblock \bibinfo{title}{Graphical models}.
\newblock \bibinfo{publisher}{Oxford University Press}.
\bibitem[{Lipton et~al.(2012)Lipton, Kim, Park, Hulkower, Gardin, Shifteh, Kim,
  Zimmerman, Lipton and Branch}]{lipton2012robust}
\bibinfo{author}{Lipton, M.L.}, \bibinfo{author}{Kim, N.},
  \bibinfo{author}{Park, Y.K.}, \bibinfo{author}{Hulkower, M.B.},
  \bibinfo{author}{Gardin, T.M.}, \bibinfo{author}{Shifteh, K.},
  \bibinfo{author}{Kim, M.}, \bibinfo{author}{Zimmerman, M.E.},
  \bibinfo{author}{Lipton, R.B.}, \bibinfo{author}{Branch, C.A.},
  \bibinfo{year}{2012}.
\newblock \bibinfo{title}{Robust detection of traumatic axonal injury in
  individual mild traumatic brain injury patients: intersubject variation,
  change over time and bidirectional changes in anisotropy}.
\newblock \bibinfo{journal}{Brain imaging and behavior} \bibinfo{volume}{6},
  \bibinfo{pages}{329--342}.
\bibitem[{Liu et~al.(2014)Liu, Lai, Wang, Hao, Chen, Zhou, Yu and
  Hong}]{liu2014MRI}
\bibinfo{author}{Liu, X.}, \bibinfo{author}{Lai, Y.}, \bibinfo{author}{Wang,
  X.}, \bibinfo{author}{Hao, C.}, \bibinfo{author}{Chen, L.},
  \bibinfo{author}{Zhou, Z.}, \bibinfo{author}{Yu, X.}, \bibinfo{author}{Hong,
  N.}, \bibinfo{year}{2014}.
\newblock \bibinfo{title}{{{A} combined {D}{T}{I} and structural {M}{R}{I}
  study in medicated-naïve chronic schizophrenia}}.
\newblock \bibinfo{journal}{Magn Reson Imaging} \bibinfo{volume}{32},
  \bibinfo{pages}{1--8}.
\bibitem[{Marion et~al.(2011)Marion, Curley, Schwab and
  Hicks}]{marion_proceedings_2011}
\bibinfo{author}{Marion, D.W.}, \bibinfo{author}{Curley, K.C.},
  \bibinfo{author}{Schwab, K.}, \bibinfo{author}{Hicks, {and} the mTBI
  Diagnostics~Wor, R.R.}, \bibinfo{year}{2011}.
\newblock \bibinfo{title}{Proceedings of the {Military} {mTBI} {Diagnostics}
  {Workshop}, {St}. {Pete} {Beach}, {August} 2010}.
\newblock \bibinfo{journal}{Journal of Neurotrauma} \bibinfo{volume}{28},
  \bibinfo{pages}{517--526}.
\newblock \URLprefix
  \url{http://www.liebertonline.com/doi/abs/10.1089/neu.2010.1638},
  \DOIprefix\doi{10.1089/neu.2010.1638}.
\bibitem[{Mayer et~al.(2014)Mayer, Bedrick, Ling, Toulouse and
  Dodd}]{mayer2014methods}
\bibinfo{author}{Mayer, A.R.}, \bibinfo{author}{Bedrick, E.J.},
  \bibinfo{author}{Ling, J.M.}, \bibinfo{author}{Toulouse, T.},
  \bibinfo{author}{Dodd, A.}, \bibinfo{year}{2014}.
\newblock \bibinfo{title}{Methods for identifying subject-specific
  abnormalities in neuroimaging data}.
\newblock \bibinfo{journal}{Human brain mapping} \bibinfo{volume}{35},
  \bibinfo{pages}{5457--5470}.
\bibitem[{Miyata et~al.(2009)Miyata, Hirao, Namiki, Fujiwara, Shimizu,
  Fukuyama, Sawamoto, Hayashi and Murai}]{miyata2009schiz}
\bibinfo{author}{Miyata, J.}, \bibinfo{author}{Hirao, K.},
  \bibinfo{author}{Namiki, C.}, \bibinfo{author}{Fujiwara, H.},
  \bibinfo{author}{Shimizu, M.}, \bibinfo{author}{Fukuyama, H.},
  \bibinfo{author}{Sawamoto, N.}, \bibinfo{author}{Hayashi, T.},
  \bibinfo{author}{Murai, T.}, \bibinfo{year}{2009}.
\newblock \bibinfo{title}{{{R}educed white matter integrity correlated with
  cortico-subcortical gray matter deficits in schizophrenia}}.
\newblock \bibinfo{journal}{Schizophr. Res.} \bibinfo{volume}{111},
  \bibinfo{pages}{78--85}.
\bibitem[{Pasternak et~al.(2014)Pasternak, Koerte, Bouix, Fredman, Sasaki,
  Mayinger, Helmer, Johnson, Holmes, Forwell, Skopelja, Shenton and
  Echlin}]{pasternak2014neuro}
\bibinfo{author}{Pasternak, O.}, \bibinfo{author}{Koerte, I.K.},
  \bibinfo{author}{Bouix, S.}, \bibinfo{author}{Fredman, E.},
  \bibinfo{author}{Sasaki, T.}, \bibinfo{author}{Mayinger, M.},
  \bibinfo{author}{Helmer, K.G.}, \bibinfo{author}{Johnson, A.M.},
  \bibinfo{author}{Holmes, J.D.}, \bibinfo{author}{Forwell, L.A.},
  \bibinfo{author}{Skopelja, E.N.}, \bibinfo{author}{Shenton, M.E.},
  \bibinfo{author}{Echlin, P.S.}, \bibinfo{year}{2014}.
\newblock \bibinfo{title}{{{H}ockey {C}oncussion {E}ducation {P}roject, {P}art
  2. {M}icrostructural white matter alterations in acutely concussed ice hockey
  players: a longitudinal free-water {M}{R}{I} study}}.
\newblock \bibinfo{journal}{J. Neurosurg.} \bibinfo{volume}{120},
  \bibinfo{pages}{873--881}.
\bibitem[{Planetta et~al.(2016)Planetta, Ofori, Pasternak, Burciu, Shukla,
  DeSimone, Okun, McFarland and Vaillancourt}]{planetta2016brain}
\bibinfo{author}{Planetta, P.J.}, \bibinfo{author}{Ofori, E.},
  \bibinfo{author}{Pasternak, O.}, \bibinfo{author}{Burciu, R.G.},
  \bibinfo{author}{Shukla, P.}, \bibinfo{author}{DeSimone, J.C.},
  \bibinfo{author}{Okun, M.S.}, \bibinfo{author}{McFarland, N.R.},
  \bibinfo{author}{Vaillancourt, D.E.}, \bibinfo{year}{2016}.
\newblock \bibinfo{title}{{{F}ree-water imaging in {P}arkinson's disease and
  atypical parkinsonism}}.
\newblock \bibinfo{journal}{Brain} \bibinfo{volume}{139},
  \bibinfo{pages}{495--508}.
\bibitem[{Ruscio(2008)}]{ruscio2008constructing}
\bibinfo{author}{Ruscio, J.}, \bibinfo{year}{2008}.
\newblock \bibinfo{title}{Constructing confidence intervals for spearman’s
  rank correlation with ordinal data: A simulation study comparing analytic and
  bootstrap methods}.
\newblock \bibinfo{journal}{Journal of Modern Applied Statistical Methods}
  \bibinfo{volume}{7}, \bibinfo{pages}{7}.
\bibitem[{Savadjiev et~al.(2014)Savadjiev, Rathi, Bouix, Smith, Schultz, Verma
  and Westin}]{savadjiev2014media}
\bibinfo{author}{Savadjiev, P.}, \bibinfo{author}{Rathi, Y.},
  \bibinfo{author}{Bouix, S.}, \bibinfo{author}{Smith, A.R.},
  \bibinfo{author}{Schultz, R.T.}, \bibinfo{author}{Verma, R.},
  \bibinfo{author}{Westin, C.F.}, \bibinfo{year}{2014}.
\newblock \bibinfo{title}{{{F}usion of white and gray matter geometry: a
  framework for investigating brain development}}.
\newblock \bibinfo{journal}{Med Image Anal} \bibinfo{volume}{18},
  \bibinfo{pages}{1349--1360}.
\bibitem[{Sun and Genton(2011)}]{sun2011functional}
\bibinfo{author}{Sun, Y.}, \bibinfo{author}{Genton, M.G.},
  \bibinfo{year}{2011}.
\newblock \bibinfo{title}{Functional boxplots}.
\newblock \bibinfo{journal}{Journal of Computational and Graphical Statistics}
  \bibinfo{volume}{20}.
\bibitem[{Vergara et~al.(2016)Vergara, Mayer, Damaraju, Kiehl and
  Calhoun}]{vergara2016neuro}
\bibinfo{author}{Vergara, V.M.}, \bibinfo{author}{Mayer, A.},
  \bibinfo{author}{Damaraju, E.}, \bibinfo{author}{Kiehl, K.},
  \bibinfo{author}{Calhoun, V.D.}, \bibinfo{year}{2016}.
\newblock \bibinfo{title}{{{D}etection of {M}ild {T}raumatic {B}rain {I}njury
  by {M}achine {L}earning {C}lassification using {R}esting {S}tate {F}unctional
  {N}etwork {C}onnectivity and {F}ractional {A}nisotropy}}.
\newblock \bibinfo{journal}{J. Neurotrauma} .
\bibitem[{White et~al.(2009)White, Schmidt and Karatekin}]{white2009white}
\bibinfo{author}{White, T.}, \bibinfo{author}{Schmidt, M.},
  \bibinfo{author}{Karatekin, C.}, \bibinfo{year}{2009}.
\newblock \bibinfo{title}{White matter ‘potholes’ in early-onset
  schizophrenia: a new approach to evaluate white matter microstructure using
  diffusion tensor imaging}.
\newblock \bibinfo{journal}{Psychiatry Research: Neuroimaging}
  \bibinfo{volume}{174}, \bibinfo{pages}{110--115}.
\bibitem[{Yoldemir et~al.(2015)Yoldemir, Ng and Abugharbieh}]{yoldemir2015ipmi}
\bibinfo{author}{Yoldemir, B.}, \bibinfo{author}{Ng, B.},
  \bibinfo{author}{Abugharbieh, R.}, \bibinfo{year}{2015}.
\newblock \bibinfo{title}{{{C}oupled {S}table {O}verlapping {R}eplicator
  {D}ynamics for {M}ultimodal {B}rain {S}ubnetwork {I}dentification}}.
\newblock \bibinfo{journal}{Inf Process Med Imaging} \bibinfo{volume}{24},
  \bibinfo{pages}{770--781}.
\bibitem[{Zhu et~al.(2012)Zhu, Li, Guo, Jiang, Zhang, Zhang, Chen, Deng,
  Faraco, Jin et~al.}]{zhu2012dicccol}
\bibinfo{author}{Zhu, D.}, \bibinfo{author}{Li, K.}, \bibinfo{author}{Guo, L.},
  \bibinfo{author}{Jiang, X.}, \bibinfo{author}{Zhang, T.},
  \bibinfo{author}{Zhang, D.}, \bibinfo{author}{Chen, H.},
  \bibinfo{author}{Deng, F.}, \bibinfo{author}{Faraco, C.},
  \bibinfo{author}{Jin, C.}, et~al., \bibinfo{year}{2012}.
\newblock \bibinfo{title}{Dicccol: dense individualized and common
  connectivity-based cortical landmarks}.
\newblock \bibinfo{journal}{Cerebral cortex} , \bibinfo{pages}{bhs072}.
\bibitem[{Zhu et~al.(2013)Zhu, Li, Jiang, Chen, Shen and
  Liu}]{zhu2013exploring}
\bibinfo{author}{Zhu, D.}, \bibinfo{author}{Li, X.}, \bibinfo{author}{Jiang,
  X.}, \bibinfo{author}{Chen, H.}, \bibinfo{author}{Shen, D.},
  \bibinfo{author}{Liu, T.}, \bibinfo{year}{2013}.
\newblock \bibinfo{title}{Exploring high-order functional interactions via
  structurally-weighted lasso models}, in: \bibinfo{booktitle}{Information
  Processing in Medical Imaging}, \bibinfo{organization}{Springer}. pp.
  \bibinfo{pages}{13--24}.

\end{thebibliography}
\end{document}